\documentstyle[preprint,aps]{revtex}

\begin{document}
\draft
\preprint{ }

\title{\bf Two-Dimensional Vortex Lattice Melting}

\author{Jun Hu and A.H. MacDonald}

\address{Department of Physics, Indiana University,
 Bloomington, Indiana 47405}

\date{\today}
\maketitle

\begin{abstract}

  We report on a Monte-Carlo study of two-dimensional
Ginzburg-Landau superconductors in a magnetic field
which finds clear evidence
for a first-order phase transition characterized by
broken translational symmetry of the superfluid density.
A key aspect of our study is the introduction of a quantity
proportional to the Fourier transform of the superfluid density
which can be sampled efficiently in Landau gauge Monte-Carlo
simulations and which satisfies a useful sum rule.
We estimate the latent heat per vortex of the melting transition
to be $\sim 0.38 k_B T_M$ where $T_M$ is the
melting temperature.

\end{abstract}


\narrowtext

  In mean-field theory type II Ginzburg-Landau
superconductors in a magnetic field have an unusual
second-order phase transition.  In the low-temperature
($T < T_c^{MF}$) phase discovered by Abrikosov\cite{abrikosov}
the zeros of the superconducting order
parameter (vortices) form a lattice and the system exhibits
both broken translational symmetry and off-diagonal
long-range order (ODLRO).
Unusual aspects of the transition are related to the
Cooper-pair Landau level structure\cite{cpll} which causes the
mean field instability of the disordered phase to
occur simultaneously at $T_{c}^{MF}$ in a macroscopic number
of channels.  However, the nature of this phase transition
is qualitatively altered by thermal fluctuations.
Interest in the effect of thermal fluctuations on
the thermodynamic properties of type II superconductors
has increased since the discovery of high-temperature
superconductors which have an unusually short coherence length
so that fluctuation effects are important over a relatively wide
temperature interval surrounding $T_c^{MF}$.

For $D$ dimensional superconductors fluctuations in a magnetic
field at temperatures
well above $T_c^{MF}$, where different channels are
independent, are like those of a $D-2$ dimensional
system\cite{leeshenoy} at zero magnetic field suggesting
that the mean-field phase transition to the Abrikosov state
will be destroyed by fluctuations for $D < 4$.
High-temperature perturbative
expansions\cite{ref:1,ref:4}, even when evaluated to high-order where
coupling between different channels becomes important,
show no evidence of a transition for $D=3$ or $D=2$
between the high-temperature fluid state and
Abrikosov's vortex-lattice state.
The results of Monte-Carlo simulations for $D=2$ have been
controversial.   Te\v sanovi\' c and Xing\cite{ref:3}
and Kato and Nagaosa\cite{kato}
find evidence for a phase transition at a
temperature below $T_c^{MF}$ while
O'Neill and Moore\cite{ref:2} have concluded that
the Abrikosov phase transition is suppressed by thermal fluctuations.
In this letter we present\cite{prelim} the results of a Monte-Carlo
simulation for $D=2$ in which we find unambiguous evidence for a
first-order phase transition.

The free energy density of a Ginzburg-Landau superconductor is
given by
\begin{equation}
f[\Psi]  = \alpha (T) \vert \Psi \vert^2 + \frac{\beta}{2} \vert \Psi
\vert^4 + \frac{1}{2m^*}\vert
(-i \hbar \nabla - 2e \vec A)\Psi \vert^2.
\label{eq:2}
\end{equation}
($ F \equiv \int d^2 \vec r  f [ \Psi ( \vec r)]$.)
The quadratic terms in Eq.~(\ref{eq:2}) are minimized by
order-parameters which correspond to a lowest Landau level
(LLL) wavefunction for the Cooper pairs.  It follows that
the mean-field theory superconducting instability occurs
at $T_c^{MF}$ ($\alpha_H(T_c^{MF})=0; \alpha_H = \alpha
+ \hbar e B / m^*$ ) for all Cooper pair states which are in the
LLL but only at much lower temperatures for
channels corresponding to higher Landau level Cooper
pair wavefunctions.  In this work we adopt the
LLL approximation in which we assume that fluctuations
in higher Landau level channels can be neglected\cite{ref:3,caveat1}
and consider only the two dimensional limit where variations
of the order parameter along the $\hat z$ direction can be
neglected.  In the LLL approximation the order parameter
is defined up to an overall scale factor by its zeros,
i.e. by the positions of the vortices. (This property has been used
by Te\v{s}anovi\'c and collaborators~\cite{ref:3,tesan92} to develop
many useful insights.)
This limit applies to films thinner than a coherence length and to
layered systems when the inter-layer coupling can be neglected.
We choose the Landau gauge ($\vec A = (0, Bx, 0)$) and
apply quasi-periodic boundary conditions to the order parameter
inside a finite system with lengths $L_x$ and $L_y$.
(For thin films, especially those formed of strongly type II
materials it is a good approximation to ignore fluctuations
in the vector potential $\vec A$.)
The order parameter $ \Psi (\vec r) $ can then be
expanded in the form,
\begin{equation}
\Psi (\vec r) =  \big({ |\alpha_H| \pi \ell^2 L_z\over \beta}\big)^{1/2}
\sum_j C_j [ \sum_s (L_yL_z)^{-1/2}(\pi
\ell^2)^{-1/4} \exp (i y X_{j,s} /\ell^2 )
\exp (-(x-X_{j,s})^2/ 4 \ell^2)]
\label{eq:1}
\end{equation}
In Eq.~(\ref{eq:1}) $X_{j,s} = j 2 \pi \ell^2/ L_y + s L_x$,
$\ell^2 = \hbar c / 2 e B$, $s $ runs over all integers and
$j$ runs from $1$ to $N_{\phi} = L_x L_y / (2 \pi \ell^2) $ which
must be chosen to be an integer.

A central role in our study is played by the
superfluid-density spatial correlation function, whose Fourier
transform is defined by
\begin{equation}
\chi_{SFD}(\vec k) \equiv
{ 1 \over L_x L_y } \int d^2 \vec r \int d^2 \vec r'
\langle |\psi(\vec r)|^2 |\psi(\vec r'|^2 \rangle
\exp [i \vec k \cdot ( \vec r - \vec r') ]
\label{eq:may1}
\end{equation}
We evaluate $\chi_{SFD}(\vec k)$ by expressing it in terms of
\begin{equation}
\Delta (\vec k) \equiv
\frac{1}{N_\phi}\sum_{j_1j_2}\bar C_{j_1} C_{j_2}
\delta_{j_2-j_1-n_y}
\exp [-ik_x (X_{j_1}+X_{j_2})/2]
\label{eq:may2}
\end{equation}
where for a finite system $\vec k = 2 \pi (n_x/L_x,n_y/L_y)$
, $\delta_j =1$ if $j$ is a multiple of $N_{\phi}$ and is
zero otherwise,
and $X_{j} \equiv X_{j,0}$. (Note that $\Delta_0 \equiv
\Delta (\vec k = 0)$ is proportional to the
integrated superfluid density.)
$\Delta(\vec k)$ is conveniently sampled in our Landau gauge
Monte Carlo simulations and
\begin{equation}
\chi_{SFD}(\vec k) = {N_{\phi}^2  \over L_x L_y }
({\alpha_H \pi \ell^2 L_z \over \beta})^2
\exp [- k^2 \ell^2 / 2 ]
\langle |\Delta (\vec k)|^2 \rangle
\label{eq:may3}
\end{equation}
Moreover $\Delta(\vec k)$ satisfies the following
sum rule for each configuration of the Ginzburg-Landau system,
\begin{equation}
{ 1 \over N_{\phi}} \sum_{\vec k}
[\vert \tilde \Delta (\vec k) \vert^2 -1/N_{\phi}] = 0
\label{equ:5}
\end{equation}
where $\tilde \Delta (\vec k) \equiv \Delta (\vec k) / \Delta_0$.
Note that $\tilde \Delta (\vec k)$ depends only on
the distribution of $| \Psi (\vec r) |^2 $ and not on
its overall magnitude.  Eq.~(\ref{equ:5}) reflects
the LLL restrictions on the superfluid density distribution.
(For a finite system, both $n_x$ and $n_y$ in the sum
over $\vec k$ in Eq.~(\ref{equ:5}) range over any $N_{\phi}$
consecutive values.)

In the vortex-liquid state $\chi_{SFD}(\vec k)$
should be a smooth function of wavevector and if
the sum over $\vec k$ in Eq.~(\ref{equ:5}) is to converge we
must have $ \lim_{|k| \to \infty} | \tilde \Delta (\vec k) |^2
\to N_{\phi}^{-1}$ and hence
that
\begin{equation}
\lim_{|k| \to \infty} \chi_{SFD}(\vec k) =
{\Delta_0^2 \alpha_H L_z \over  2 \beta}
\exp [- k^2 \ell^2 / 2 ]
\label{eq:may4}
\end{equation}
It is readily verified\cite{junhupt} that Eq.~(\ref{eq:may4})
is satisfied for all $\vec k \ne 0$ when $ T \gg T_c^{MF}$
and the vortex fluid is completely uncorrelated.
On the other hand, in a vortex-lattice state
$ \Delta(\vec k) = \Delta_0 \delta_{\vec k, \vec G}$ where
$\vec G$ is a reciprocal lattice vector.  To see that
Eq.~(\ref{equ:5}) is satisfied in this case
note that there are $N_{\phi}$
wavevectors per Brioullin zone in the Abrikosov state.
Eq.~(\ref{equ:5}) tells us that averages of
$|\tilde \Delta (\vec k)|^2$ over large areas of reciprocal
space yield $N_{\phi}^{-1}$ irrespective of the degree of
correlation among the vortices.
$N_{\phi} \langle | \tilde \Delta (\vec k)|^2 \rangle  -1$ provides a
very convenient measure of the degree of vortex
correlation in a system.

We can express the Ginzburg-Landau free energy in terms of $\vert
\Delta (\vec k) \vert^2 $ as follows
\begin{equation}
\int \frac{f[\Psi]}{k_B T} d\vec r \equiv
\frac{E_{\beta}(\beta,\Delta_0)}{k_B T}
= N_{\phi} g^2 [\rm{sgn}(\alpha_H) \Delta_0
 +  {\beta[\tilde \Delta]  \Delta_0^2 \over 4}]
\label{equ:6}
\end{equation}
where
\begin{equation}
\beta[\tilde \Delta] \equiv
\sum_{\vec k}\vert \tilde \Delta (\vec k) \vert^2 \exp [-\frac{k^2l^2}{2}].
\label{equ:thurs}
\end{equation}
and $g \equiv  \alpha_H (\pi \ell^2 L_z/\beta K_B T)^{1/2}$.
$\beta[\tilde \Delta]$ has its minimum value in the
Abrikosov state and increases as the vortex positions become
less correlated.
It is readily verified that in the uncorrelated vortex
fluid $\beta[\tilde \Delta] = 2 $ while for the triangular
lattice Abrikosov state $\beta[\tilde \Delta] = \beta_A \sim 1.159595.$
(This relatively weak variation in $\beta$ was exploited recently
\cite{tesan92} by Te\v sanovi\' c {\it et al.}.)
We regard $\beta[\tilde \Delta]$ and $\Delta_0$ as the two
intensive thermodynamic variables which characterize the state of the
LLL Ginzburg-Landau system.  We can define an entropy which measures
the function-space volume associated with a given $\beta [\tilde
\Delta ]$ and $\Delta_0$ by $S_{\beta}(\beta,\Delta_0) \equiv
k_B  \ln ( W(\beta,\Delta_0))$ where
\begin{equation}
W(\beta,\Delta_0) \equiv  \big({ |\alpha_H| \pi \ell^2 L_z\over
\beta}\big)^{N_{\phi}}
 \prod_{j} \int d\bar C_j dC_j  \delta ( \beta - \beta[\tilde \Delta])
 \delta ( \Delta_0 - \sum_j \bar C_j C_j)
\label{new1}
\end{equation}
With this definition the free energy, $\beta$ and $\Delta_0$
at any value of $g$ can be determined by minimizing
\begin{equation}
F_{\beta}(\beta,\Delta_0) \equiv E_{\beta}(\beta,\Delta_0) - T
S(\beta,\Delta_0)
\label{new2}
\end{equation}
with respect to $\beta$ and $\Delta_0$. ($F_{\beta}$ is extensive
so fluctuations become negligible in the thermodynamic limit.)
We will use Eq.~(\ref{new2}) to interpret the Monte Carlo results
discussed below.

Using the Metropolis algorithm we have determined distribution
functions for several quantities\cite{tobepub} including
$E_\beta$, $\Delta_0$, $\beta[\tilde \Delta]$, and $ |\tilde \Delta
(\vec k)|^2 $ as a function of both $g$ and $N_{\phi}$.
Finite system shapes have been
chosen to accommodate perfect triangular lattices.
For all simulations the order parameter was
initialized to the Abrikosov lattice value
and the first $10^4$ Monte Carlo steps were discarded.
Some typical results for $ \langle \vert \tilde \Delta (\vec k)
\vert^2 \rangle $ at $T < T_c^{MF}$ are shown in Fig.~(\ref{fig1}).
At $g^2 =30$ the vortex fluid has developed strong
correlations.  For $N_{\phi}=120$,
$N_{\phi} \langle |\tilde \Delta (\vec G)|^2 \rangle \sim 3$
which is three times larger than for the high-temperature
uncorrelated flux-fluid but still $\sim  40 $ times
smaller than its mean field value.  For $g^2 = 50$,
$\langle |\tilde \Delta (\vec G) |^2 \rangle $
has increased to more than half its mean field value.
The insets in Fig.~(\ref{fig1}) show the dependence of $| \tilde \Delta
( \vec G)|^2 $ on system size for these two values of $g^2$.
For $g^2 = 30$,  $\langle |\tilde \Delta (\vec G) |^2 \rangle \sim
N_{\phi}^{-1.0} $  as expected in the fluid state while for
$g^2 = 50$, $\langle |\tilde \Delta (\vec G) |^2 \rangle \sim
N_{\phi}^{-0.13}$, consistent
with the quasi-long-range order expected in the Abrikosov state.
Fig.~(\ref{fig2}) shows that for a given system size
$\langle |\tilde \Delta (\vec G) |^2 \rangle $ increases relatively
abruptly at $g^2 \sim 42.5$ suggesting the occurrence of a
phase transition.

To examine this possibility and to
determine the order of the phase transition
we have examined the dependence of the energy
distribution function\cite{ref:5,ref:6} on
system size for $g^2 \sim 43$ and $N_\phi = 80, 100, 120, 144,
168 $. The results are shown in Fig.~(\ref{fig3}).
For each systems size the number of Monte Carlo
steps required to determine these distribution functions accurately
exceeded $8\times 10^6 $.
For $N_{\phi} > 100$ a double peak structure indicative
of a first-order phase transition is clearly visible.  For
each $N_{\phi}$ the adjusted\cite{ref:5,ref:6} distribution function
at the value of $g^2$ where the peaks have equal height is plotted.
By extrapolating these values of $g^2$ to $N_{\phi} = \infty$
as shown in the inset we estimate that a first order phase
transition occurs at $g^2 = g_M^2 = 43.5 \pm 1.0$.  By comparing the
separations between the peak positions we estimate that the
latent heat per flux quantum associated with the transition is
$\sim 0.01 k_B T g_M^2 /\beta_A \sim  0.38 k_B T $.
In Fig. (3b) we compares the
$\langle |\tilde \Delta (\vec G) |^2 \rangle $ distribution
from values of the order parameter with high-energies
with that from low-energies for $N_{\phi} = 168$.
For high-energy configurations the
$\langle |\tilde \Delta (\vec G) |^2 \rangle $
is $\sim 5.0 N_{\phi}^{-1}$ while for the
low-energy configurations the distribution is peaked at $\sim 0.5$
demonstrating that the phase transition occurs between a high-energy
strongly correlated vortex fluid state
and a low-energy Abrikosov state.

In Fig. (4a) we show distribution functions for $\beta(\tilde \Delta)$
at several fixed values of $\Delta_0$ and in
Fig. (4b) we show distribution functions for $\Delta_0$ at several
fixed values of $\beta(\tilde \Delta)$ for $\beta(\tilde \Delta)$
and $\Delta_0$ near the values at which the phase transition takes
place.
The $\beta(\tilde \Delta)$ distribution function
is proportional to $\exp [(S_{\beta}(\beta, \Delta_0)/K_B -
\beta N_{\phi} g^2 \Delta_0^2 / 4] $.  (See Eq.~(\ref{new2}).
At extrema of the distribution $\partial S_{\beta} / \partial \beta =
K_B^{-1}N_{\phi} g^2 \Delta_0^2/4$.)  The double peak structure apparent
in the $\beta(\tilde \Delta)$ distribution demonstrates that the
phase transition is driven by the $S_{\beta}$ term
which describes the dependence of the volume
in order parameter space on the degree of correlation in
vortex positions.  No similar double peak structure is seen
in the $\Delta_0$ distribution confirming that the
phase transition is associated primarily with spatial correlations in
vortex positions and hence in the superfluid density rather
than with changes in the magnitude of the superconducting order
parameter.

This work was supported by the Midwest
Superconductivity Consortium through D.O.E.
grant no. DE-FG-02-90ER45427.  The authors are grateful to
D.P. Arovas and Kieran Mullen for help getting started and
to S.M. Girvin for frequent useful advice.  AHM
acknowledges instructive conversations with D.A. Huse,
D.R. Nelson, and Zlatko Te\v sanovi\' c.

\newpage

\begin{figure}
\caption{$\langle | \tilde \Delta (\vec q_x) |^2 \rangle $
at $g^2 = 30$ (a) and $g^2 = 50$ (b) for a finite system with
$N_{\phi} = 120$. ($ q_y = 0$ and $T  < T_c^{MF}$.)
The insets show the dependences of
$\langle |\tilde \Delta (\vec G) |^2 \rangle $ on system size
at these $g^2$ values.
The dashed lines in the inset of (a) is proportional to
$N_\phi^{-1.0}$ while that in
the inset of (b) is proportional to $N_\phi^{-0.13}$.
($\vec G$ is a member of the
first shell of reciprocal lattice vectors of the
Abrikosov lattice.)  These averages were obtained
from $1 \sim 2 \times 10^6 $ Monte-Carlo steps.}
\label{fig1}
\end{figure}

\begin{figure}
\caption{Dependence of $\langle |\tilde \Delta (\vec q) |^2 \rangle$
on $g^2$ at $N_{\phi}=120$ for
$\vec q = \vec G$ and for $\vec q \ne \vec G$ where $\vec G$
is a reciprocal lattice vector of the Abrikosov lattice.}
\label{fig2}
\end{figure}

\begin{figure}
\caption{(a): Landau-Ginzburg energy distribution function at
the finite system phase transition point for various system sizes.
Energies are in units of the mean-field condensation energy,
$N_{\phi} k_B T g^2/ \beta_A$.
The ratio of the peak heights to the intermediate minimum grows with
system size but for the sizes we are able to study does not
yet show the $\exp [ c N_{\phi}^{1/2}] $ behavior
expected at large $N_{\phi}$.
The inset shows the dependence of the $g^2$ at the
phase transition on system size.
(b): low-energy and high-energy
cuts of the distribution function for $|\tilde \Delta (\vec G)|^2$.}
\label{fig3}
\end{figure}

\begin{figure}
\caption{Distribution functions of (a) $\beta[\tilde \Delta]$ for
several values of $\Delta_0$ and of (b) $\Delta_0$ for several values
of $\beta[\tilde \Delta]$.}
\label{fig4}
\end{figure}

\end{document}